\documentclass[preprint]{revtex4-1}

\usepackage{graphicx}

\begin{document}

\title{Performance Evaluation of Klystron Beam Focusing System with Anisotropic Ferrite Magnet}


\author{Yasuhiro Fuwa}
\affiliation{Department of Physics, Kyoto University, Kyoto, Kyoto 606-8502, Japan \email{fuwa@kyticr.kuicr.kyoto-u.ac.jp}}

\author{Yoshihisa Iwashita}
\affiliation{Institute for Chemical Research, Kyoto University, Uji, Kyoto 611-0011, Japan}


\begin{abstract}%
A klystron beam focusing system using permanent magnets, which increases reliability in comparison with electromagnet focusing system, is reported. A prototype model has been designed and fabricated for a 1.3 GHz, 800 kW klystron for evaluation of the feasibility of the focusing system with permanent magnets.  In order to decrease the production cost and to mitigate complex tuning processes of the magnetic field, anisotropic ferrite magnet is adopted as the magnetic material.  As the result of a power test, 798 $\pm$ 8 kW peak output power was successfully achieved with the prototype focusing system.  Considering a power consumption of the electromagnet focusing system, the required wall-plug power to produce nominal 800 kW output power with the permanent magnet system is less than that with electromagnet.  However, the power conversion efficiency of the klystron with the permanent magnet system was found to be limited by transverse multipole magnetic fields.  By decreasing transverse multipole magnetic field components, especially the dipole and the quadrupole, the power conversion efficiency would approach to that with electromagnets. 
\end{abstract}


\maketitle

\section{Introduction}
\label{}

   The International Linear Collider (ILC) is an electron-positron collider for high-energy physics with a center-of-mass energy up to 500 GeV.  In order to achieve the final energy, the ILC should be equipped with more than sixteen thousands of accelerating cavities to accelerate electrons and positrons [1]. Order of one thousand klystrons are also needed to feed the Radio Frequency (RF) power to the cavities.  For example, the Distributed RF Scheme (DRFS), one of the RF power feed scheme formerly proposed for ILC, utilizes eight thousand midium-power klystrons [2].  Due to the large number of klystron units, even a low failure rate of each component could increase the load of maintenance work and limit availability of the facility.  

  For DRFS, the reliability of klystron itself is estimated to be fairly high because of its relatively-low output power (800 kW per klystron).  In addition to the high reliability of klystron itself, that of the whole system is also important.  The klystron beam focusing system is one of the key components in the whole system.  In fact, according to KEK injector operating statistics during 10 years in 2000s, the failures in focusing solenoids were the most frequent reasons to exchange klystron assemblies among the other reasons [3].  
  
Focusing systems with permanent magnets for klystron had been developed in some accelerator laboratories.  In the SLAC 2-mile linear accelerator [4] and the Photon Factory linac in KEK [5], the focusing systems with Alnico magnets were developed and utilized in accelerator operation.  However, both laboratories have discontinued them and have employed solenoid coils as the klystron beam focusing system.  The main reason for the discontinuances was the complexity in the field tuning.  The magnetic fields generated by the magnets in these laboratories were tuned by partial magnetization and/or demagnetization of the Alnico magnets or attaching magnetic material shims such as iron bars.  Because the tuning methods had poor reproducibilities due to irreversibility caused by the low coercivity of Alnico, the tuning process had some difficulties which took a long time.
  
  Because the number of magnet systems in ILC klystron is quite large, they must be cost effective and have ease of tuning.  This paper reports on developments of a prototype focusing magnet with ferrite magnets for ILC DRFS klystron and the result of performance evaluation for the focusing magnet by way of both experimental and numerical studies.  It should be noted that the Technical Design Report (TDR) for the ILC adopts the Distributed Klystron Scheme (DKS) using the Multi-Beam Klystron (MBK) instead of the DRFS mainly because of the higher efficiency after discussions during the GDE (Global Design Effort) for ILC.
  
  The main purpose of this study is to prove that focusing systems with permanent magnets can replace electromagnets with a similar performances (power conversion efficiency and peak power).  This paper consists of six sections as follows.  In Sec. 2,  the conceptual design of the magnet is briefly presented.  Sec. 3 describes design process of prototype magnet and the result of performance estimation using simulation.  In Sec. 4, the result of power test with prototype magnet is presented.  In Sect. 5, the effect of asymmetric fields and error fields on output power is discussed.  In Sect. 6, a brief summery is presented.

\section{Conceptual Design of Focusing System}
  In this section, the conceptual design of our focusing system with permanent magnets is briefly described.  A detailed description can be found in Ref. 6.  The anisotropic ferrite is chosen as magnetic material because of its inexpensive cost, enough remanent field ($\sim$ 4 kG), and coercivity.  Because the required magnetic field in DRFS klystron is up to 1 kG, the ferrite magnets can generate magnetic field with the sufficient magnitude.
  
  The focusing field is uni-directional along the beam axis in contrast to the Periodic Permanent Magnets (PPM) configuration [7].  Then the field distribution can be the same as the field generated by a solenoid coil except for the reverse fields in the exterior regions.  The uni-directional field distribution can avoid the damage on the klystron tube wall due to a beam loss [8] caused by the energy stop-band on the electron beam transmission.  In order to generate the required magnetic field with less amount of ferrite material, magnets are configured in a quasi-Halbach dipole configuration [9].  The configuration also contributes to reduce the stray magnetic field outside the system and to prevent field distortion due to some disturbance caused by magnetic materials in vicinity.  The image of the magnet configuration and the magnetic field flux distribution calculated by PANDIRA [10] are shown in Fig. 1.  The magnets are divided into a number of segments and each segment is movable independently.  By virtue of this segmentation and movability, magnets can be set near the klystron tube after the insertion of the tube to the focusing alcove.  The placement of the magnets near the tube can reduce the required volume of magnets to generate focusing field.  The local tuning of the magnetic field can be also realized by fine positionings of the segments.  This tuning scheme has a good reproducibility because of the high coercivity of ferrite magnets and is a much easier process compared with the early studies where the field distributions were tuned by the cumbersome tuning methods as stated in the earlier section [5].

\begin{figure}[!tbh]
    \centering
    \includegraphics[width=100mm]{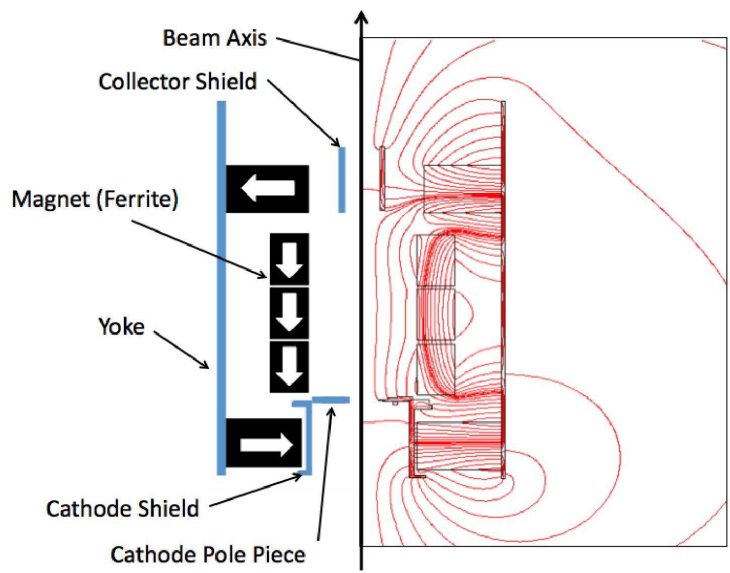}
    \caption{Configuration of magnets (left side of the figure). The calculated field distribution is shown in the right half part.}
    \label{f1}
\end{figure}

\section{Design of Prototype Model}
 For evaluation of the performance of the focusing system described in Sect. 2, a prototype was designed for DRFS klystrons.  The klystron assumed to be used in DRFS was a E37501 [11] manufactured by Toshiba Electron Tube \& Devices Co., Ltd.  A modulating anode imbedded in the klystron makes the radius of the high-voltage insulator around the cathode part large, which requires the large outer radius of the oil tank, and hence, inflates the magnet volume for that part .  RADIA 4.29 [12] was used for the magnetic field design.  After the fabrication cost estimation, the basic dimension of the ferrite piece was chosen as 150 mm $\times$ 100 mm cross section and 25.4 mm thickness.  The optimized result of the magnet and the yoke configuration is shown in Fig. 2.  The properties of each magnet are described in Ref. 6.  The magnets are designed so as to avoid interference with the cooling pipes and the RF input port on the klystron outer wall.  The fields in cathode region were shielded by the oil tank whose cylinder is made of ferromagnetic iron.  By moving the positions of the magnets, the field distribution in the beam drift region can be adjusted.  All magnets can evacuate from the central region to make the space to insert the klystron from the top.  The tuned field distribution for efficient beam focusing is shown in Fig. 3.  In this figure, the normalized value of the magnetic field distribution is shown instead of the absolute value due to a non-disclosure agreement between authors and the tube supplier.  In the discussion afterward, the positions of cathode, input cavity, and output cavity are indicated by $z_{cathode}$, $z_{input}$, and $z_{output}$, respectively.  Compared with the field generated by electromagnet (EM) in Fig. 3, the field by permanent magnet (PM) has ripples in the beam drift region.  These ripples were caused by longitudinal segmentation of the magnets.  For estimation of the effect on beam focusing, beam envelope was calculated as cylindrically symmetric problems by DGUN code [13].  Comparing the beam envelopes for PM and EM (Fig. 4), the difference between two envelopes is not significant, which can conclude that such ripples can be neglected for a klystron design.

  \begin{figure}[!tbh]
    \centering
    \includegraphics[width=90mm]{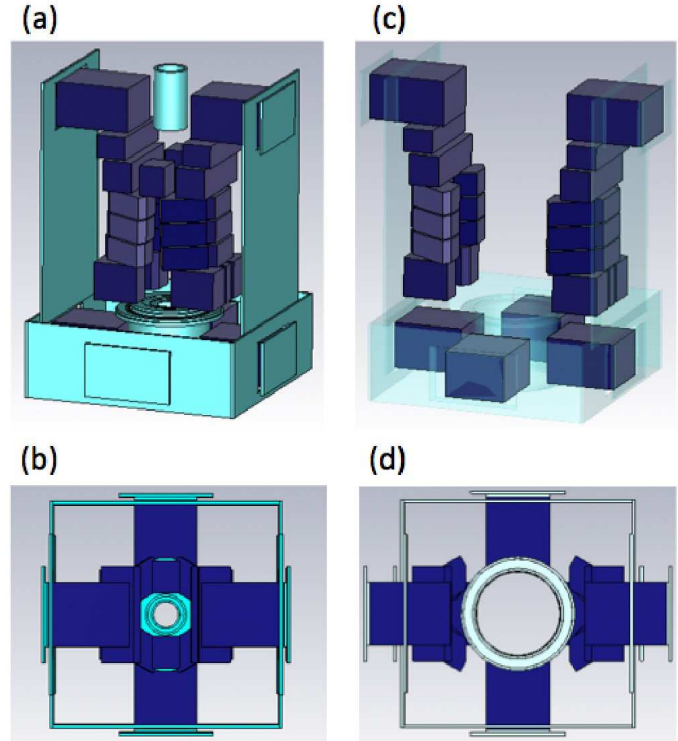}
    \caption{Designed configuration of magnets and iron yoke.  (Left) Magnet position for beam operation. (Right) Magnet position for klystron installation. The objects in (b) and (d) are seen from upper side.}
    \label{f2}
\end{figure}

\begin{figure}[!tbh]
    \centering
    \includegraphics[width=80mm]{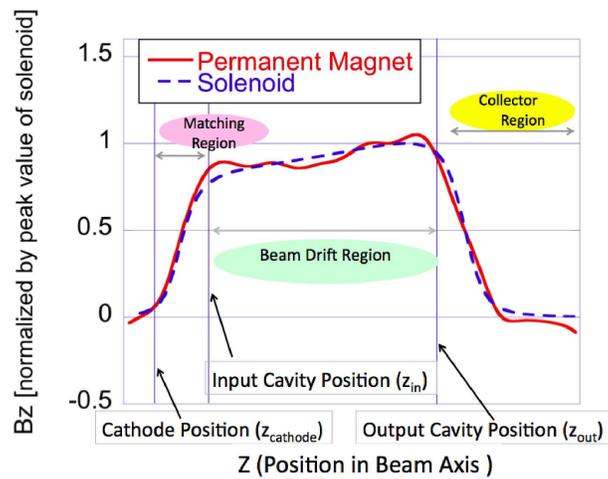}
    \caption{Magnetic field distribution along the beam axis.  The electron beam from the cathode is accelerated and matched to the Brillouin flow optics in the matching region and is transported through beam drift region.  It finally goes into the collector region to be damped.}
    \label{f3}
\end{figure}

\begin{figure}[!tbh]
    \centering
    \includegraphics[width=75mm]{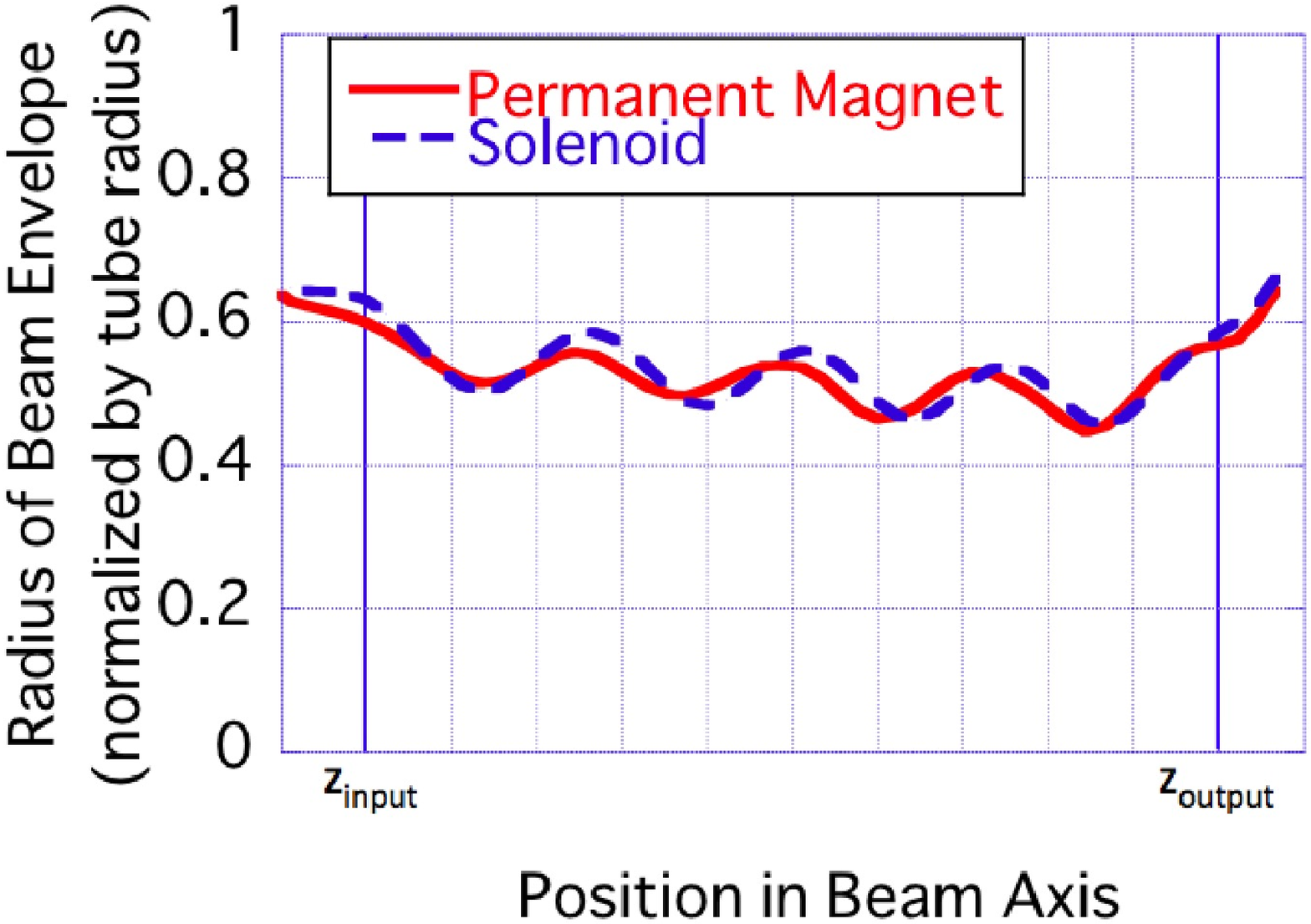}
    \caption{Beam envelopes calculated by 2.5 dimensional simulation with DGUN.}
    \label{f4}
\end{figure}
  
  Multipole components of magnetic fields in the transverse planes are also generated because the distribution of magnets is not azimuthally symmetric.  The dominant component is quadrupole and the strength of the quadrupole component reaches 20 G/cm (see Fig. 12).  Higher order components such as an octupole component are also existing, but their magnitude are less than one tenth of the quadrupole component in the beam region.  Quadrupole fields deform a circular beam into an elliptical beam [14].  In the case of our magnet system, electron beams are magnetically confined and rotate around the center of beam (Brillouin flow).  Therefore, the effect of the quadrupole field is not simple.  For evaluation of the electron distribution in the beam propagation from the cathode to the collector, 3D beam tracking simulation was performed with CST Particle studio [15].  Fig. 5 shows calculated beam profiles at several points along beam axis.  From the results, while the beam profile became elliptical during beam propagation in the quadrupole component field, all the particles reached to the collector without hitting the tube wall.  Beam simulations with RF modulation were also performed with PIC (Particle-in-Cell) simulation and no beam loss was observed in any beam slices.

\begin{figure}[!tbh]
    \centering
    \includegraphics[width=110mm]{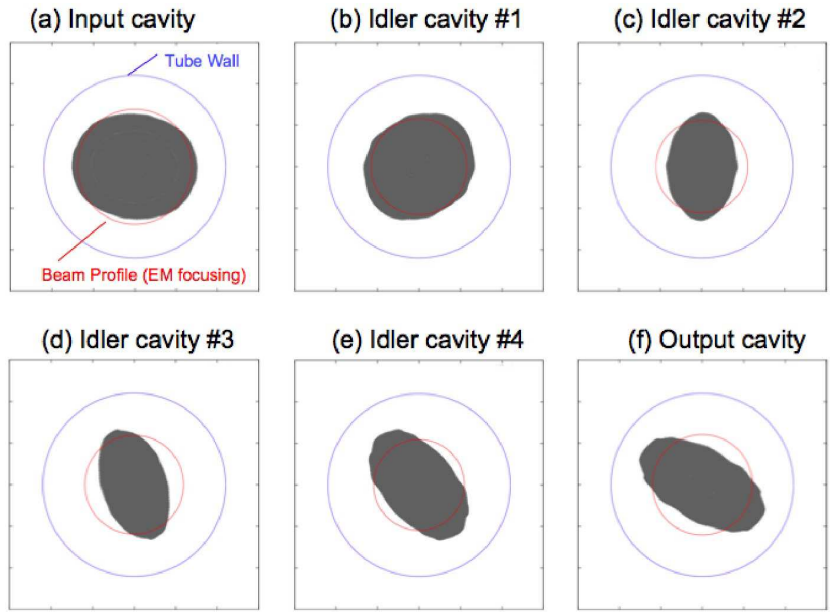}
    \caption{Beam profiles at cavity positions calculated by 3 dimensional simulation with CST particle studio.}
    \label{f5}
\end{figure}

\section{Test with Prototype Magnet}
\subsection{Fabrication of Prototype Magnet}
Based on the design described in Sect. 3, the prototype magnet was fabricated.  The magnet pieces shaped from 150mm $\times$ 100mm $\times$ 25.4mm ferrite ingots to designed shapes were glued together by acrylic cured resin.  Then, each glued segment was also glued on an iron plate or an aluminum block.  All the magnets were supported by these metallic structures.  The pictures of the fabricated prototype are shown in Fig. 6.  The magnets could be moved as designed for the klystron insertion and the field tuning.  The typical accuracy of the magnet position alignment is less than 1 mm.  

In field tuning process, the magnetic field distributions \{$B_x(z)$, $B_y(z)$, $B_z(z)$ \} on the beam axis were measured with 3-axis Hall probe (Senis 3MH3).  After some iteration processes of tuning by adjusting magnet positions, the longitudinal field distribution could be adjusted to the designed one within 2 \% error in the beam drift region.  The field in the matching region (see Fig. 3) was more precisely adjusted to suppress ripples of the beam envelope.  The typical transverse field distribution is shown in Fig. 7.  The transverse fields were caused by residual positioning errors of magnets in the tuning process, the mechanical errors of magnet shapes, and magnetization errors in the fabrication process.  Careful field tuning can make this transverse field less than 10 G (corresponding to 10 \% magnitude compared with the longitudinal magnetic field).  This magnitude (fraction) of transverse magnetic field on the beam axis was comparable to that of the focusing system developed at SLAC [16].  

The displacement of the beam center can be estimated with an assumption that electrons follow the magnetic flux lines.
The displacement of the flux line $\Delta r$  at the position $z$ can be estimated by integrating the flux line angle (${B_{transverse}}/{B_z}$) from the beam axis.  Using these facts, the beam displacements in $x$, $y$ directions can be derived in a first order approximation as follows.

The $x$-component of the transverse field can be described as
\begin{equation}
B_{transverse,x}(x,y,z) = B_x(z) +\frac{x}{r} B_r(r,z),
\end{equation}
where $r = \sqrt{x^2+y^2}$ is displacement from the beam axis, $B_x(z)$ is the $x$-component of the dipole field on the axis, and $B_r(r,z)$ is azimuthally symmetric field component at the distance $r$ from the axis.  Using the condition div$B=0$, the $B_r(r,z)$ can be derived from the longitudinal field component on the axis as
\begin{equation}
B_r(r, z) = - \frac{1}{2} \frac{\partial B_{z}(z)}{\partial z} r.
\end{equation}
Hereby, the displacement of beam center $\Delta r$ can be calculated as
\begin{equation}
\Delta x(z) = \int^z_{z_{cathode}} \frac{B_{x} (\xi) + \frac{\Delta x(\xi)}{\Delta r(\xi)} B_r(\Delta r(\xi), \xi)}{B_z(\xi)} d\xi,
\end{equation}

\begin{equation}
\Delta y(z) = \int^z_{z_{cathode}} \frac{B_{y} (\xi) + \frac{\Delta y(\xi)}{\Delta r(\xi)} B_r(\Delta r(\xi), \xi)}{B_z(\xi)} d\xi,
\end{equation}

\begin{equation}
\Delta r(z) = \sqrt{ \Delta x(z)^2 + \Delta y(z)^2},
\end{equation}
where $\Delta x$ and $\Delta y$ are the displacement of beam center in $x$ and $y$ direction, respectively.  The calculated result of the displacement of beam center due to the transverse magnetic field estimated by equations (2) - (5) is shown in Fig. 8 together with a result from CST particle studio. Although the space charge effects are neglected in the first order estimation, the estimated beam displacement gives a rough guess close to the result with CST, which reduce a lot of efforts and computation time.  The reason of the discrepancy should be non-linear effect due to space charges and beam-wall interactions.

\begin{figure}[!tbh]
    \centering
    \includegraphics[width=60mm]{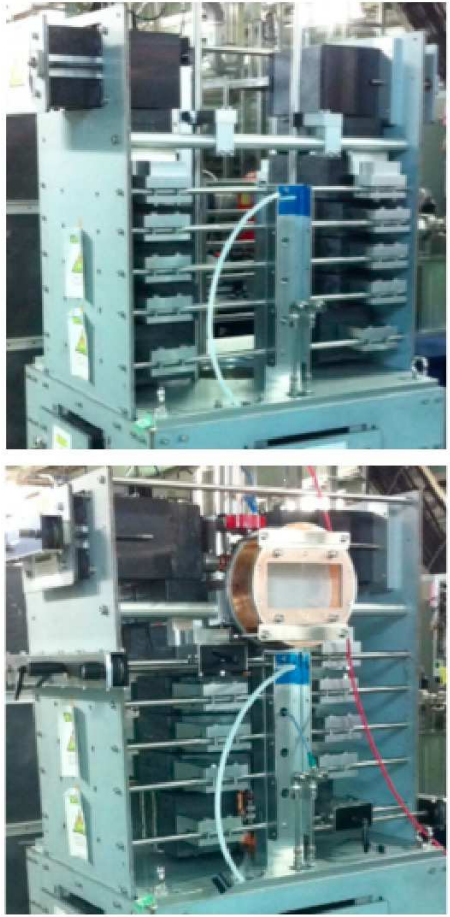}
    \caption{Fabricated prototype magnet. (Top) Klystron insertion mode: magnets are retracted towards the outer sides and making the space to let the cathode assembly part go through, which is the most thick part of the klystron. (Bottom) klystron operating mode; magnets are moved to designed position to generate the proper focusing field.}
    \label{f4}
\end{figure}  

\begin{figure}[!tbh]
    \centering
    \includegraphics[width=75mm]{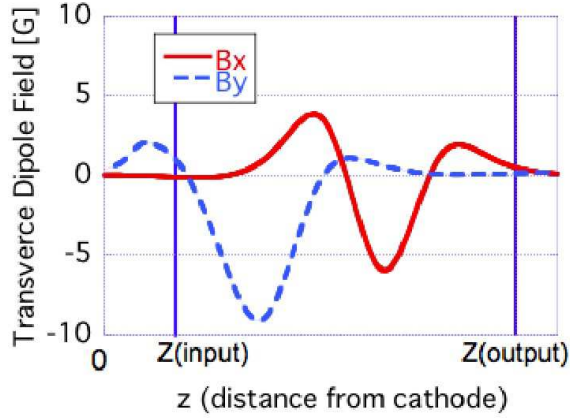}
    \caption{Typical measured transverse (dipole) magnetic fields along the beam axis.}
    \label{f4}
\end{figure}

\begin{figure}[!tbh]
    \centering
    \includegraphics[width=75mm]{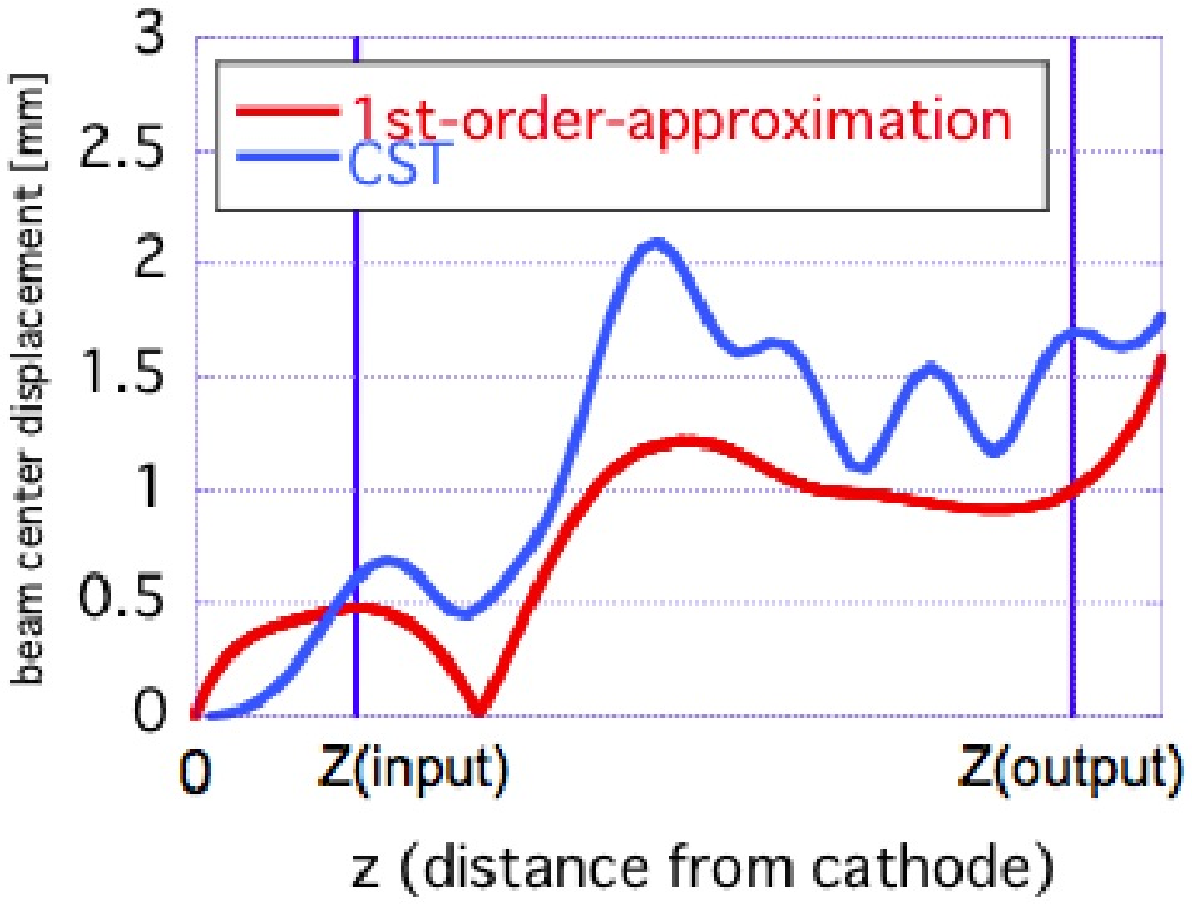}
    \caption{The estimated beam-center displacement due to the transverse magnetic field (see Fig. 7).  Both the calculated result of the 1st order approximation and the result from a massive simulation with CST particle studio are presented.}
    \label{f4}
\end{figure}
  
\subsection{Klystron Power Test}
Klystron power test was performed in KEK STF.  A 1 MW dummy load was 
connected.  A modulator at STF building in KEK supplied the powers for the klystron, which includes a heater power, a cathode high voltage and a modulating anode pulse.  A preliminary result of the power test was presented in Ref. 11.  In the initial stage of experiment, a diode test was carried out.  In this test, no RF power was fed to the input cavity and the beam was just transported to the collector.  The cathode voltage was set between 45 kV and 67.5 kV, and the modulating anode voltage at beam extraction was 0.26 times cathode voltage.  The measured perveance of extracted beam was 1.29 $\pm$ 0.03 $[\mu \mathrm{A}/\mathrm{V}^{3/2}]$.  This value is the same as the result using the solenoid (1.29 $\pm$ 0.04 $[\mu \mathrm{A}/\mathrm{V}^{3/2}]$).  During the beam extraction test, the temperature rise of the cooling water for klystron body was monitored, and no  significant temperature rise was detected.  These results suggested that the amount of the electron loss on the klystron tube body was ignorable and our prototype magnets had sufficient beam transport efficiency.

Then, the test with RF excitation was carried out.  RF input-output characterization was measured at several cathode voltages.  The result is shown in Fig. 9.  At 65.8 kV cathode voltage, the saturated output power reached 798$\pm8$  kW.  Because the required output power per klystron for DRFS is 800 kW, the system with permanent magnet focusing had enough capability to produce the RF power.

\begin{figure}[!tbh]
    \centering
    \includegraphics[width=75mm]{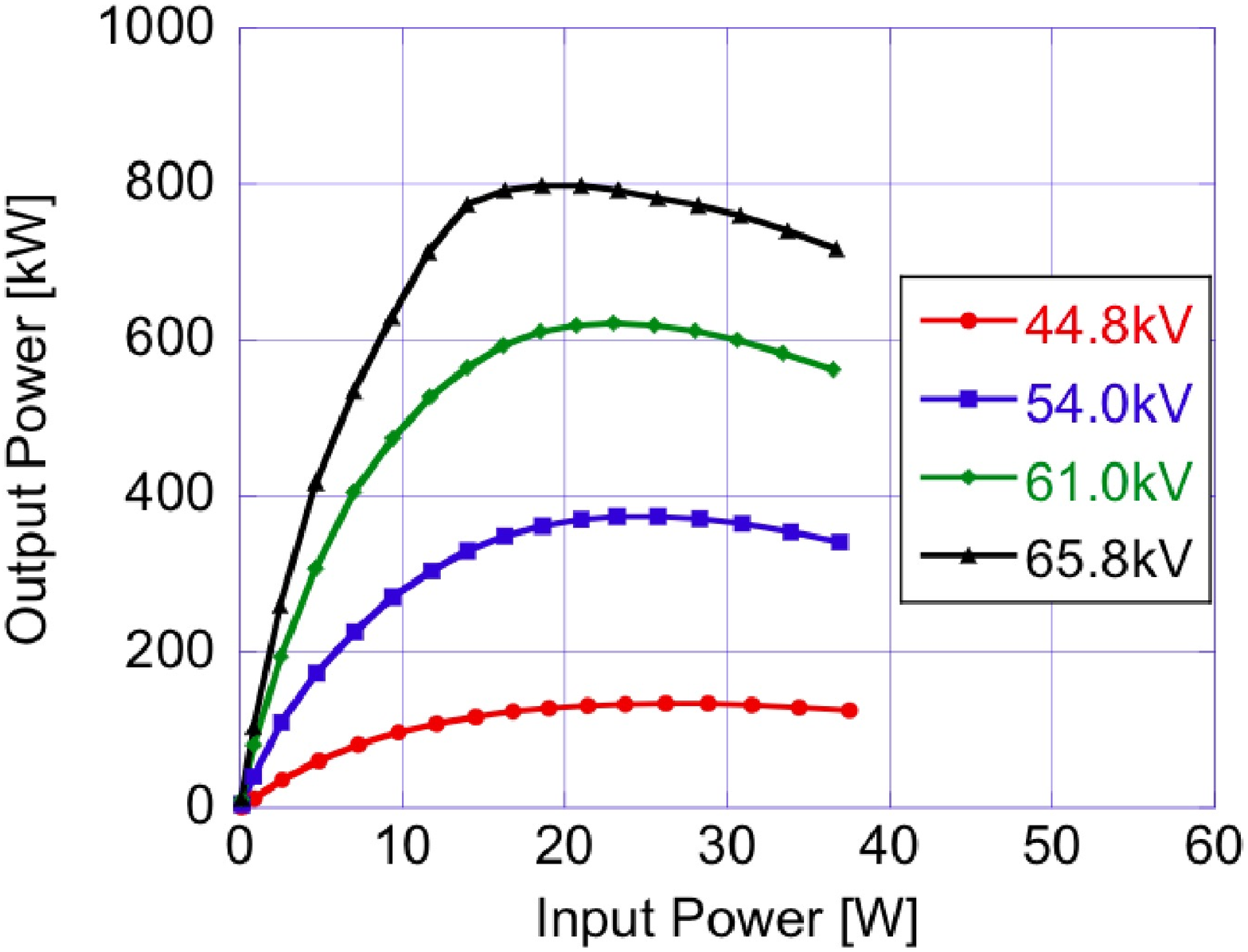}
    \caption{The output RF power of the DRFS klystron with permanent magnet focusing as function of input RF power at four cathode voltages.}
    \label{f4}
\end{figure}

\section{Discussion}
\subsection{Output power compared with solenoid focusing}

For comparison, the saturated output power measurement was also performed with electromagnet (EM).  The results for both PM and EM are shown in Fig. 11.  The output power with the EM is slightly larger than that with PM by about 10 \%.  For production of 5.6 kW RF average power (nominal parameter for DRFS; 800 kW peak RF power with 5 Hz repetition rate and 1.5 ms pulse duration), the required average beam power is about 9.6 kW (59 \% efficiency for beam-to-RF energy conversion) with the solenoid magnet and 10.5 kW (54 \% efficiency) with the prototype permanent magnet.  Because the solenoid magnet continuously consumes 1.2 kW power, the total power consumption of the prototype PM case exceeds the same level of the EM case.  This is a good result considering the limited experiment time duration of a few weeks.  The power conversion efficiency of a klystron with an improved PM focusing system would be enhanced at least as the same as the EM case and required power from wall plug would be reduced.  In DRFS, the number of klystrons is about 8,000, and the total eliminated power in the total facility would be 9.6 MW.

\begin{figure}[!tbh]
    \centering
    \includegraphics[width=75mm]{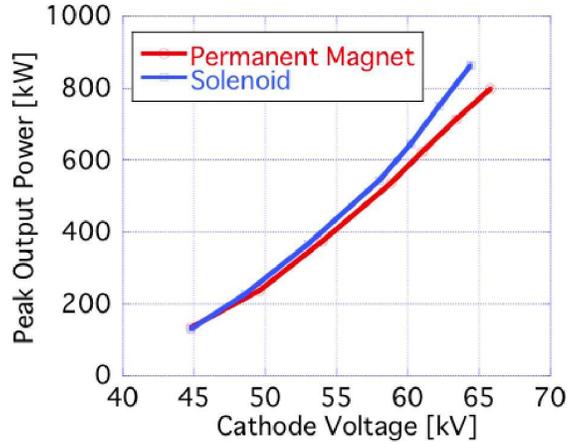}
    \caption{The results of the power test.  Output RF power is shown as functions of cathode voltage with permanent magnet focusing and solenoid focusing.}
    \label{f4}
\end{figure}

  The main reason of the discrepancy of output power between the PM case and the EM case is considered to be the beam deformation and displacement due to the small multipole field components.  The beam deformation and displacement lead the asymmetric distribution of the particles and may cause excitation of higher order modes in the cavities such as TM110 and TM210 modes.  The kick due to the higher order modes and dipole magnetic field might cause beam loss and decrease output power.  However, a significant temperature rise of the tube outer wall was not observed, indicating that there were no beam losses in the klystron tube during the power test.  The deformation of the beam decreases clearance between the beam envelope and the tube wall (including cavity iris), and pushes up the global coupling factors between the beam and the cavities, which represents a coupling factor averaged over all the electrons in the beam (see Appendix).
  
The calculated value of the global coupling factor $M$ focused by the PM without dipole error was about 1 \% larger than the EM case.  As the beams propagate through the tube, the beams with the larger energy modulation would result the larger density modulation and the larger RF current of the beam.  Since the cavities in the klystron are designed and tuned for the beam focused with the EM focusing, the electron beam with larger beam-cavity coupling factor tends to be longitudinally over-focused at the output cavity.  According to the 1D klystron simulation, while a small increase less than 1.7 \% relative to the global coupling factor value for the EM case lead to increase of the output power, larger increase of the global coupling factor causes decrease of the output power.  Therefore, with perturbation on the global coupling factor larger than 1.7 \%,  the maximum bunching efficiency and the induced RF current in the output cavity decrease compared with the EM focusing.  With the PM case with dipole error, the perturbation of the global coupling factor was larger than 3 \% and expected output power is less than the EM case.

For the evaluation of the effects of the beam deformations, PIC (Particle in Cell) simulations with CST particle studio were also performed.  The simulations with the PM focusing were performed with two conditions; 1) without transverse dipole field (without imperfections of the magnets), 2) with  the transverse error field.  In the simulation, the excited voltages of higher order modes were less than those of the fundamental mode by two orders of magnitude.  In addition, any beam loss in the beam drift region was observed in the simulations.  The simulation results showed the tendency of changing the output power with the PM focusing compared with the EM focusing as supposed by 1D simulation discussed above.  The calculated output power with the PM focusing (without dipole field) was 874 $\pm$ 6 kW, while that with the EM focusing was 865 $\pm$ 5 kW.  Therefore, the results suggest that the quadrupole fields reduce the output power.  With the dipole field shown in Fig. 7, the peak output power was calculated as 809 $\pm$ 7 kW (the output power was decreased to 95 $\pm$ 2 \% of that with the EM focusing).  The reduction of the output power is supposed to be caused by the enhancement of the beam-cavity coupling factor due to the beam center displacement.  In Fig. 11, the input-output characteristics both measured in the power test and calculated by the PIC simulations are presented.  In the PIC simulations, the induced voltages of idler cavities ($V_{cavity}$) were also monitored (see Table 1).  The calculated results shows that all the induced voltages with the PM focusing are larger than those with the EM focusing, which indicates that the former case has a larger global coupling factors than the latter.  

\begin{figure}[!tbh]
    \centering
    \includegraphics[width=75mm]{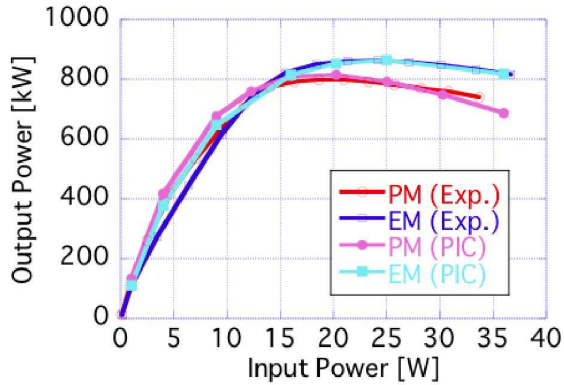}
    \caption{The input-output characteristic both measured in the power test and calculated with PIC (Particle-in-Cell) simulation.}
    \label{f4}
\end{figure}

\begin{table*}[h]
  \begin{center}
   \caption{Calculated voltages induced in the idler cavities with 20 W input power.}
  \begin{tabular}{c||cc|c}
    Idler cavity \# & $V_{cavity}$(PM) [kV] & $V_{cavity}$(EM) [kV] & Ratio[ $V_{cavity}$(PM) / $V_{cavity}$(EM)] [\%]\\ \hline
    1 & 19.56  & 19.27  & 101.3 \\
    2 & 6.86  & 6.84  & 100.3 \\
    3 & 20.04  & 19.75  & 101.4 \\
    4 & 34.93  & 33.83  & 103.2 \\
  \end{tabular}
  \end{center}
\end{table*}

\subsection{Suppression of the transverse field}
As discussed in the previous subsection, the beam-to-RF power conversion efficiency was supposed to be deteriorated by the transverse multipole magnetic field on the klystron designed for the axi-symmetric field focusing.  After the completion of an allowed experiment time in the busy STF schedule, improvements on the focusing system was studied.  For the recovery of the output power, the multipole field components have to be suppressed.  The transversal segmentation of the magnets with a higher symmetry would result less multipole components.  The magnet configuration with the two-fold axial symmetry about beam axis generates the quadrupole components, which was supposed to have not much effect on the efficiency at the design stage.  It should be noted that the construction of a more highly symmetric structure leads to more difficulty in the magnet position alignment and the complexity of the mechanical support of magnets.  Considering the above mentioned situation, and the fabrication and the assembling cost, a transverse field suppressor is favorable rather than the modification of the magnet structure in a fancy way.

Although the multipole field suppressor can be realized either by active methods, such as correction coils, or by passive methods such as additional magnetic materials, the passive methods have the advantage of higher reliability.  Passive field suppression can be achieved by installing a transverse field filter with anisotropic macroscopic permeability, where the relative permeability in the longitudinal direction is close to unity and that in transverse directions are sufficiently larger than 1.  A filter with such an anisotropic permeability can be realized by sparsely stacked rings made of soft-magnetic material such as silicon steel sheets.  For our permanent magnet system, the inner radii of the rings can touch the klystron outer wall (60 mm) and the outer radii are 70 mm, while the thickness of the rings is 0.5 mm.  For evaluation of the suppression capability of the transverse field, magnetic fields were calculated with CST EM studio.  From the calculated results with various ring intervals, the ring train with 12 mm spacing could suppress the multipole field component most efficiently without a significant reduction of longitudinal magnetic field component.  With 12 mm interval rings, the quadrupole field is suppressed down to less than 1/5 (see Fig. 12).  The calculated transverse dipole field was also reduced to less than 1/2 of the field without the filter.  While the transverse multipole field components were suppressed efficiently, the effect of installing the rings on the longitudinal field components around the axis was less than 1 \% ($<$ 0.5 mm / 12 mm).  

PIC simulation results with the filter showed the improvement of the output power.  The calculated output power with the filter-installed PM focusing was 875 $\pm$ 5 kW.  The output power become almost the same as that with the EM focusing case.  A PIC simulation including the magnet imperfections with field filter showed no output power reduction.  These results indicate that such a filter effectively reduces the transverse multipole components.

\begin{figure}[!tbh]
    \centering
    \includegraphics[width=75mm]{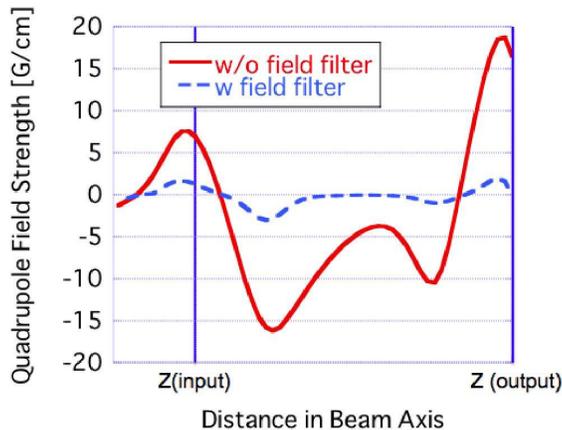}
    \caption{Transverse quadrupole magnetic field calculated with CST EM studio. The transverse field filter reduce the quadrupole field to less than 20 \% of the field without the filter.}
    \label{f4}
\end{figure}





\section{Conclusion}
Klystron beam focusing systems with permanent magnets are well suited to the large accelerator facilities such as future linear colliders to enhance their reliability and reduce costs.  Even the prototype system with the ferrite magnets for 1.3 GHz 800 kW klystrons exhibits performance almost the same as with EM focusing.  This PM system could not be optimized because of the limited trial and the experiment time.  For reduction of the fabrication costs, anisotropic ferrite was adopted as magnetic material and the magnets were designed to be pushed-in after the klystron tube insertion to the focusing alcove.  The movable magnet feature could also be usable for field tuning process.  The technologies adopted in the prototype system can be applied for other klystrons.  Especially, the application of the technology to the Multi Beam Klystron would be important for ILC.  Furthermore, by mitigating the effects of the transverse magnetic field, the focusing system with permanent magnets can save not only the power consumption in the accelerator facilities, but also push up the reliability of the whole system by eliminating the possible water leakages from the solenoid coils and the possible failures of power supplies.  

\section*{Acknowledgements}
This work was supported by the Collaborative Research Program of Institute for Chemical Research, Kyoto University (grant \# 2016-10).  The authors thank Mr. Y. Okubo at Toshiba Electron Tubes \& Devices Co., Ltd for giving us the klystron data and for the fruitful discussions on the klystron, and thank Dr. H. Hayano at KEK for his continuous encouragement and fruitful comments.  The authors also thank Dr. S. Fukuda, Dr. S. Michizono, and Dr. T. Matsumoto at KEK for their support on executing the klystron power test.  The experimental results presented in this paper could not be aquired without their help.


%

\appendix

\section{Beam-Cavity Coupling Factor}
  The coupling factor between a single particle and a cavity has the radius dependency through the transit time factor
  
\begin{equation}
T(r) = \frac{\int^{g/2}_{-g/2} E_z(r, z) \cos \bigl({\frac{2 \pi z}{\beta \lambda} \bigr)dz}}{\int^{g/2}_{-g/2} E_z(r, z) dz},
\end{equation}
where $g$, $\beta$ and $\lambda$ are the length of the cavity gap, the electron velocity relative to the speed of light and the wave length at the klystron operating frequency, respectively.  In our study (electron energy of about 65 kV at the klystron operating frequency 1.3 GHz), $\beta \lambda$ is 0.11 m.  $E_z(r, z)$ is the $z$ component of the electric field for the fundamental cavity mode.  Since the radial component of electron velocity can be neglected compared with the longitudinal component, $r$ can be treated as a constant in the integration.  The global beam-cavity coupling factor $M$ can be evaluated as the average of $T(r)$ for all electrons in the beam.  Using the beam-cavity coupling factor $M$, the relation between the amplitude of the RF current in the cavity ($i_{cavity}$) and that of the beam ($i_{beam}$) can be written as
\begin{equation}
i_{cavity} = M \  i_{beam}.
\end{equation}
The voltage induced in the cavity ($V_{cavity}$) is denoted as
\begin{equation}
V_{cavity} = R \  i_{cavity},
\end{equation}
where $R$ is the shunt impedance of the cavity.  The energy modulation $\Delta U$  by the induced voltage is described as
\begin{equation}
\Delta U = T(r) \  V_{cavity} \cos{\phi},
\end{equation}
where $r$ and $\phi$ are the position from beam axis and RF phase at the time when the electron arrives at the cavity gap.  Since the T(r) is a monotonically increasing function of $r$ for the re-entrant cavity, large $r$ leads increasing of $M$ and the modulation gain enhancement.  
\end{document}